\documentclass[twocolumn, twocolappendix]{aastex631}
\usepackage{amsmath}

\begin{document}

\title{An unbiased method of measuring the ratio of two data sets}
\shorttitle{An unbiased method of measuring the ratio of two data sets}
\shortauthors{Sun et al.}

\author[0000-0001-8771-306X]{Zeyang Sun}
\altaffiliation{zeyangsun@sjtu.edu.cn}
\affiliation{Department of Astronomy, School of Physics and Astronomy, Shanghai Jiao Tong University, Shanghai, 200240, China}
\affiliation{Key Laboratory for Particle Astrophysics and Cosmology (MOE) / Shanghai Key Laboratory for Particle Physics and Cosmology, China}

\author{Pengjie Zhang}
\altaffiliation{zhangpj@sjtu.edu.cn}
\affiliation{Department of Astronomy, School of Physics and Astronomy, Shanghai Jiao Tong University, Shanghai, 200240, China}
\affiliation{Key Laboratory for Particle Astrophysics and Cosmology (MOE) / Shanghai Key Laboratory for Particle Physics and Cosmology, China}
\affiliation{Division of Astronomy and Astrophysics, Tsung-Dao Lee Institute, Shanghai Jiao Tong University, Shanghai, 200240, China}

\author[0000-0003-0296-0841]{Fuyu Dong}
\affiliation{South-Western Institute for Astronomy Research, Yunnan University, Kunming, 650500, China}

\author[0000-0002-7336-2796]{Ji Yao}
\affiliation{Shanghai Astronomical Observatory (SHAO), Nandan Road 80, Shanghai, 200030, China}

\author[0000-0001-8534-837X]{Huanyuan Shan}
\affiliation{Shanghai Astronomical Observatory (SHAO), Nandan Road 80, Shanghai, 200030, China}
\affiliation{University of Chinese Academy of Sciences, Beijing, 100049, China}

\author[0000-0002-9253-053X]{Eric Jullo}
\affiliation{Aix-Marseille Univ, CNRS, CNES, LAM, Marseille, France}

\author{Jean-Paul Kneib}
\affiliation{Aix-Marseille Univ, CNRS, CNES, LAM, Marseille, France}
\affiliation{Institute of Physics, Laboratory of Astrophysics, Ecole Polytechnique Fédérale de Lausanne (EPFL), Observatoire de Sauverny, 1290 Versoix, Switzerland}

\author{Boyan Yin}
\affiliation{Department of Physics, Carnegie Mellon University, Pittsburgh, Pennsylvania 15312, USA}


\begin{abstract}
In certain cases of astronomical data analysis, the meaningful physical quantity to extract is the ratio $R$ between two data sets. Examples include the lensing ratio, the interloper rate in spectroscopic redshift samples, the decay rate of gravitational potential and $E_G$ to test gravity. However, simply taking the ratio of the two data sets is biased, since it renders (even statistical) errors in the denominator into systematic errors in $R$. Furthermore, it is not optimal in minimizing statistical errors of $R$. Based on Bayesian analysis and the usual assumption of Gaussian error in the data, we derive an analytical expression of the posterior PDF $P(R)$. This result enables fast and unbiased $R$ measurement, with minimal statistical errors. Furthermore, it relies on no underlying model other than the proportionality relation between the two data sets. Even more generally, it applies to the cases where the proportionality relation holds for the underlying physics/statistics instead of the two data sets directly. It also applies to the case of multiple ratios ($R\rightarrow {\bf R}=(R_1,R_2,\cdots)$). We take the lensing ratio as an example to demonstrate our method. We take lenses as DESI imaging survey galaxies, and sources as DECaLS cosmic shear and \emph{Planck} CMB lensing. We restrict the analysis to the ratio between CMB lensing and cosmic shear. The resulting $P(R)$, for multiple lens-shear pairs, are all nearly Gaussian. The S/N of measured $R$ ranges from $4.9$ to $8.4$.  We perform several tests to verify the robustness of the above result.   

\end{abstract}

\keywords{gravitational lensing: weak --- methods: statistical --- large-scale structure of Universe --- cosmology: observations --- cosmic background radiation}


\section{Introduction} \label{sec:intro}

In astronomy data analysis (e.g. cosmological analysis), we often need to combine different data sets for joint analysis. In certain cases, the desirable quantity to extract from the joint data is the ratio of two data sets.  We list several examples as follows. (1) The lensing ratio \citep{Jain2003, Bernstein2004, ZhangJun2005}. This is the ratio of galaxy-galaxy lensing of different sources (e.g. cosmic shear at various redshifts and CMB lensing) but identical lenses. The ratio provides a clean measure on the geometry of the universe. It has been measured for various data combinations \citep{Taylor2007, Das2009, Kitching2015, Baxter2016, Miyatake2017, Prat2018, Prat2019, DESY1_2018, DESY3_2021}. (2) The decay rate $DR$ of gravitational potential, which is the ratio between galaxy-ISW cross-correlation and galaxy-lensing cross-correlation \citep{Zhang2006ISW}. It has recently been measured by \citet{Dong2022} combining DESI galaxy-Planck ISW/lensing cross-correlations. The $DR$ measurement, in combination with BAO or SNe Ia data, improved constraints of dark energy by $\sim 20$-$50\%$ \citep{Dong2022}.  (3) Interloper rate due to line confusion in spectroscopic redshift surveys.  This particular error in redshift measurement can be approximated as the ratio between the cross-correlation of two target galaxy samples and the auto-correlation (e.g. \citet{Addison2019, Farrow2021, Gong2021}). (4) $E_G$ as a probe of gravity at cosmological scales \citep{Zhang2007E_G}. It is essentially the ratio between cross-correlations of galaxy-velocity and  galaxy-lensing. It has been measured by   \citet{Reyes2010, Leonard2015, Blake2016, Pullen2016, Alam2017, delaTorre2017, Amon2018, Singh2019, Skara2020S, ZhangYucheng2021}, using various data of redshift space distortion (RSD) and weak lensing.  

The ratio is therefore straightforward to measure by simply taking the ratio of the two corresponding measurements. However, there are a number of reasons to improve this naive estimator. (1) The naive estimator is not only sub-optimal in terms of statistical errors, but also biased. Suppose that we have two data sets, $D_1=\bar{D}_1(1+n_1)$ and $D_2=\bar{D}_2(1+n_2)$. Here $\bar{D}$ is the true value the theory predicts $\bar{D}_2/\bar{D}_1=R$. $n$ is the fractional error and for brevity we assume only statistical error here ($\langle n\rangle=0$). The naive estimator $\hat{R}=D_2/D_1$ is biased since
\begin{equation}
    \begin{aligned}
        \langle \hat{R}\rangle & =\left\langle \frac{D_2}{D_1}\right\rangle=R\left(1+\langle n_1^2\rangle - \langle n_1n_2 \rangle+\cdots\right)\neq R\ .
    \end{aligned}
\end{equation}
(2) Furthermore, in some applications the physically meaningful $R$ is not directly the ratio of two data sets, but rather the ratio of some underlying models. For example, one data set is the galaxy-tangential shear cross-correlation, and the other is the galaxy-CMB lensing convergence cross-correlation. The first is related to the galaxy-lensing power spectrum through the Bessel function $J_2(x=\ell \theta)$. The second is related to $J_0(x)$ instead. So although the underlying galaxy-lensing power spectra follow the proportionality relation, the data sets do not. Another example is $E_G$, as measured by combining a 3D galaxy-velocity power spectrum inferred from RSD and a 2D galaxy-lensing angular power spectrum. In both cases, we can not simply take the ratio of two data sets to obtain the true ratio. (3) A further issue is that there are multiple $R$ of interest (namely ${\bf R}=(R_1, R_2, \cdots)$), but the corresponding data sets to measure them are correlated. This is the case for the interloper rate, and also the more general case of photo-z outliers. 

We present a likelihood-based optimal estimator of the ratio, free of the above problems. Under the usual assumption of Gaussian errors in the data, we derive the exact analytical expression for the posterior PDF $P(R)$ (or the joint PDF $P({\bf R})=P(R_1,R_2,\cdots)$). Since the expression is exact, it enables unbiased $R$ estimation with minimum statistical uncertainty. The method has been applied in a companion paper \citep{Dong2022} to measure the decay rate of gravitational potential at cosmological scales. Here we present a thorough description of the method, and further demonstrate its applicability with the lensing ratio measurement as an example. 

The paper is organized as follows. In \S \ref{sec: methodology}, we introduce the method. In \S \ref{sec: application}, we show the application of this method by measuring lensing ratios, including the measurements of the lensing ratios (\S \ref{subsec: measurements of lensing ratios}) and the consistency checks (\S \ref{subsec: consistency checks}). We conclude in \S \ref{sec: conclusions}. Appendix \ref{sec: data pre-processing} describes the data pre-processing. Appendix \ref{sec: statistics} shows the details on the lensing ratio statistics, ratio modeling, and implications of the measured ratio.

\section{Methodology} 
\label{sec: methodology}

The problem to be solved can be formulated as follows. We have two data sets, $\mathbf{d}_1$ and $\mathbf{d}_2$. The theory expectation of $\mathbf{d}_1$ is fixed by the theory parameter vector $\boldsymbol{\lambda}$, $\mathbf{d}_1 = \mathbf{A}_1\boldsymbol{\lambda} + \mathbf{n}_1$. Here the mapping matrix $\mathbf{A}_1$ of dimension $N_\mathbf{d} \times N_{\boldsymbol{\lambda}}$.  
$n_1$ is the corresponding noise, which we assume to be Gaussian with zero mean. $\mathbf{d}_2$ is fixed by the same set of theory parameter $\boldsymbol{\lambda}$ and an extra parameter $R$.  The dependence on $R$ is through the second mapping matrix $\mathbf{A}_2(R)$, $\mathbf{d}_2 = \mathbf{A}_2(R)\boldsymbol{\lambda} + \mathbf{n}_2$. The simplest case is ${\bf A}_2=R{\bf A}_1$. This is the case for  the shear ratio, where $d_i=w_i(\theta)$, $\boldsymbol{\lambda}=\langle w_1(\theta)\rangle$, and $w_i(\theta)$ are the two galaxy-shear cross-correlations. But in general $A_2$ and $A_1$ are independent, and ${\bf A}_2(R)$ can be an arbitrary function of $R$ or even ${\bf R}=(R_1,R_2,\cdots)$. For brevity, we work on the case of $R$. The extension to the more general case of ${\bf R}$ is straightforward. 

We can study this problem based on Bayesian analysis, 
\begin{equation}\label{eq: P_R}
    P(R|\mathbf{d}_1, \mathbf{d}_2) \propto \int P(\mathbf{d}_1, \mathbf{d}_2 |R, \boldsymbol{\lambda}) P(\boldsymbol{\lambda}) P_{\rm prior}(R) d\boldsymbol{\lambda} \ .
\end{equation}
We take a flat prior of $\boldsymbol{\lambda}$ ($P(\boldsymbol{\lambda}) \propto$ const.) in order not to introduce extra model dependence. $P_{\rm prior}(R)$ refers to the prior of $R$, and we explain its choice at the end of this section. What we find is that, for Gaussian distribution of the data, the marginalization over $\boldsymbol{\lambda}$ can be done analytically. So we can obtain an analytical expression of $P(R)$. The expression depends on whether ${\bf d}_{1,2}$ are correlated. We first derive the result for the simpler case of uncorrelated ${\bf d}_{1,2}$, and then proceed to the general case of correlated ${\bf d}_{1,2}$. 

\subsection{Uncorrelated \texorpdfstring{$\mathbf{d}_{1,2}$}{}}
When measurement errors in $\mathbf{d}_{1}, \mathbf{d}_2$ are uncorrelated, the joint likelihood on the right-hand-side of Eq.~\ref{eq: P_R} can be separated into the product of two individual likelihood functions,
\begin{equation}
\label{eqn:P12}
    P(\mathbf{d}_1, \mathbf{d}_2| R, \boldsymbol{\lambda}) = P(\mathbf{d}_1 | R,\boldsymbol{\lambda}) P(\mathbf{d}_2 | R,\boldsymbol{\lambda}) \ .
\end{equation}
For Gaussian distributed $\mathbf{d}_{1,2}$,
\begin{equation}
    P(\mathbf{d}_{i} | R,\boldsymbol{\lambda}) = \frac{1}{\sqrt{(2\pi)^N {\rm det} \mathbf{C}_i}} {\rm exp}\bigg[-\frac{1}{2} \mathbf{\Delta}_i^T \mathbf{C}_i^{-1} \mathbf{\Delta}_i \bigg] \ . \nonumber
\end{equation}
Here $\mathbf{\Delta}_i\equiv \mathbf{d}_i- \mathbf{A}_i\boldsymbol{\lambda}$. $\mathbf{C}_i$ is the covariance matrix of ${\bf d}_i$, ${\bf C}_i\equiv \langle \mathbf{n}_i \mathbf{n}_i^T \rangle$. All vectors are column vectors by default, like $\boldsymbol{\lambda}^T = (\lambda_1, \lambda_2, ..., \lambda_N)$. $N$ is the size of $\mathbf{d}_{1,2}$. Plugging the above expression into Eq. \ref{eq: P_R} \& \ref{eqn:P12}, 
\begin{equation}
\label{eqn:PE}
    P(R|\mathbf{d}_1, \mathbf{d}_2)\propto  \int \exp(E)P_{\rm prior}(R)d\boldsymbol{\lambda}\ .
\end{equation}
We have to ignore the proportionality prefactors which do not depend on $\boldsymbol{\lambda}$. 
$E$ in the exponential is 
\begin{equation}
\label{eq:expan exp}
\begin{aligned}
 E= -\frac{1}{2}& (\boldsymbol{\lambda}^T \mathbf{A}_1^T \mathbf{C}_1^{-1} \mathbf{A}_1 \boldsymbol{\lambda} - \boldsymbol{\lambda}^T\mathbf{A}_1^T \mathbf{C}_1^{-1}\mathbf{d}_1 - \mathbf{d}_1^T\mathbf{C}_1^{-1}\mathbf{A}_1\boldsymbol{\lambda} \\ & + \mathbf{d}_1^T \mathbf{C}_1^{-1}\mathbf{d}_1 +\boldsymbol{\lambda}^T \mathbf{A}_2^T \mathbf{C}_2^{-1} \mathbf{A}_2 \boldsymbol{\lambda} - \boldsymbol{\lambda}^T\mathbf{A}_2^T \mathbf{C}_2^{-1}\mathbf{d}_2 \\ & - \mathbf{d}_2^T\mathbf{C}_2^{-1}\mathbf{A}_2\boldsymbol{\lambda} + \mathbf{d}_2^T \mathbf{C}_2^{-1}\mathbf{d}_2) \ . 
\end{aligned}
\end{equation}
If we let $\mathbf{Q} \equiv \mathbf{Q}_1 + \mathbf{Q}_2$, ${\bf Q}_i\equiv \mathbf{A}_i^T \mathbf{C}_i^{-1} \mathbf{A}_i$, $\mathbf{T}\equiv \mathbf{T}_1 + \mathbf{T}_2$,${\bf T}_i\equiv \mathbf{A}_i^T \mathbf{C}_i^{-1}\mathbf{d}_i$ and group terms in powers of $\boldsymbol{\lambda}$, we can rewrite Eq.~\ref{eq:expan exp} as
\begin{equation} 
\label{eq:A5}
 E=   -\frac{1}{2} (\boldsymbol{\lambda}^T \mathbf{Q}\boldsymbol{\lambda} -\boldsymbol{\lambda}^T \mathbf{T} - \mathbf{T}^T \boldsymbol{\lambda}) \ .
\end{equation}
Here we ignore the two terms $\mathbf{d}_1^T \mathbf{C}_1^{-1}\mathbf{d}_1$ and $\mathbf{d}_2^T \mathbf{C}_2^{-1}\mathbf{d}_2$. These do not depend on $\boldsymbol{\lambda}$ and therefore do not affect the  shape of $P(R)$.  Since ${\bf C}^T_i={\bf C}$, $\mathbf{Q}_i^T = \mathbf{Q}_i$ and ${\bf Q}^T={\bf Q}$. This means that ${\bf Q}$ is a Hermitian matrix, with real eigenvalues. So we make the assumption that it is invertible and change the form of $E$ as 
\begin{equation}
E= -\frac{1}{2} \Big[ (\boldsymbol{\lambda} - \mathbf{Q}^{-1} \mathbf{T})^T \mathbf{Q} (\boldsymbol{\lambda} - \mathbf{Q}^{-1} \mathbf{T}) - \mathbf{T}^T \mathbf{Q}^{-1}\mathbf{T} \Big] \ .
\end{equation}
Plugging it into Eq. \ref{eqn:PE} and integrating away $\boldsymbol{\lambda}$, we obtain the first major result of this paper, 
\begin{equation}
\label{eq: P_R uncorr}
    P(R|\mathbf{d}_1, \mathbf{d}_2) \propto \left({\rm det}\mathbf{Q}\right )^{-1/2} {\rm exp} \bigg[ \frac{1}{2} \mathbf{T}^T \mathbf{Q}^{-1}\mathbf{T} \bigg] P_{\rm prior}(R) \ .
\end{equation}
Here we have used the relation
\begin{equation}
    G(\mathbf{Q}) \equiv \int {\rm exp} \bigg[ -\frac{1}{2} \mathbf{z}^T\mathbf{Q}\mathbf{z} \bigg] \prod \limits_{i=0}^{N} dz_i = (2\pi)^{N/2} ({\rm det}\mathbf{Q})^{-1/2} \ .
\end{equation}

Notice that in Eq. \ref{eq: P_R uncorr}, both ${\bf Q}$ and ${\bf T}$ depend on $R$, through ${\bf A}_2(R)$. There is no analytical expression for the best-fit $R$, so we have to numerically evaluate $P(R)$. For the numerical evaluation, we should instead evaluate
\begin{equation}
\begin{aligned}
     \ln P(R|{\bf d}_1,{\bf d}_2) = &  -\frac{1}{2} \ln \big ({\rm det}\mathbf{Q}(R) \big ) + \ln P_{\rm prior}(R) \\ &  + \frac{1}{2} \mathbf{T}^T(R) \mathbf{Q}^{-1}(R)\mathbf{T}(R)  + {\rm const.}  \ .
\end{aligned}
\end{equation}
To avoid numerical errors associated with too large/small exponential terms, a safer way is to evaluate the r.h.s. as a function of $R$,  find the maximum, and then subtract this maximum before evaluating $P(R|{\bf d}_1,{\bf d}_2)$.

\begin{figure*}[ht!]
\plotone{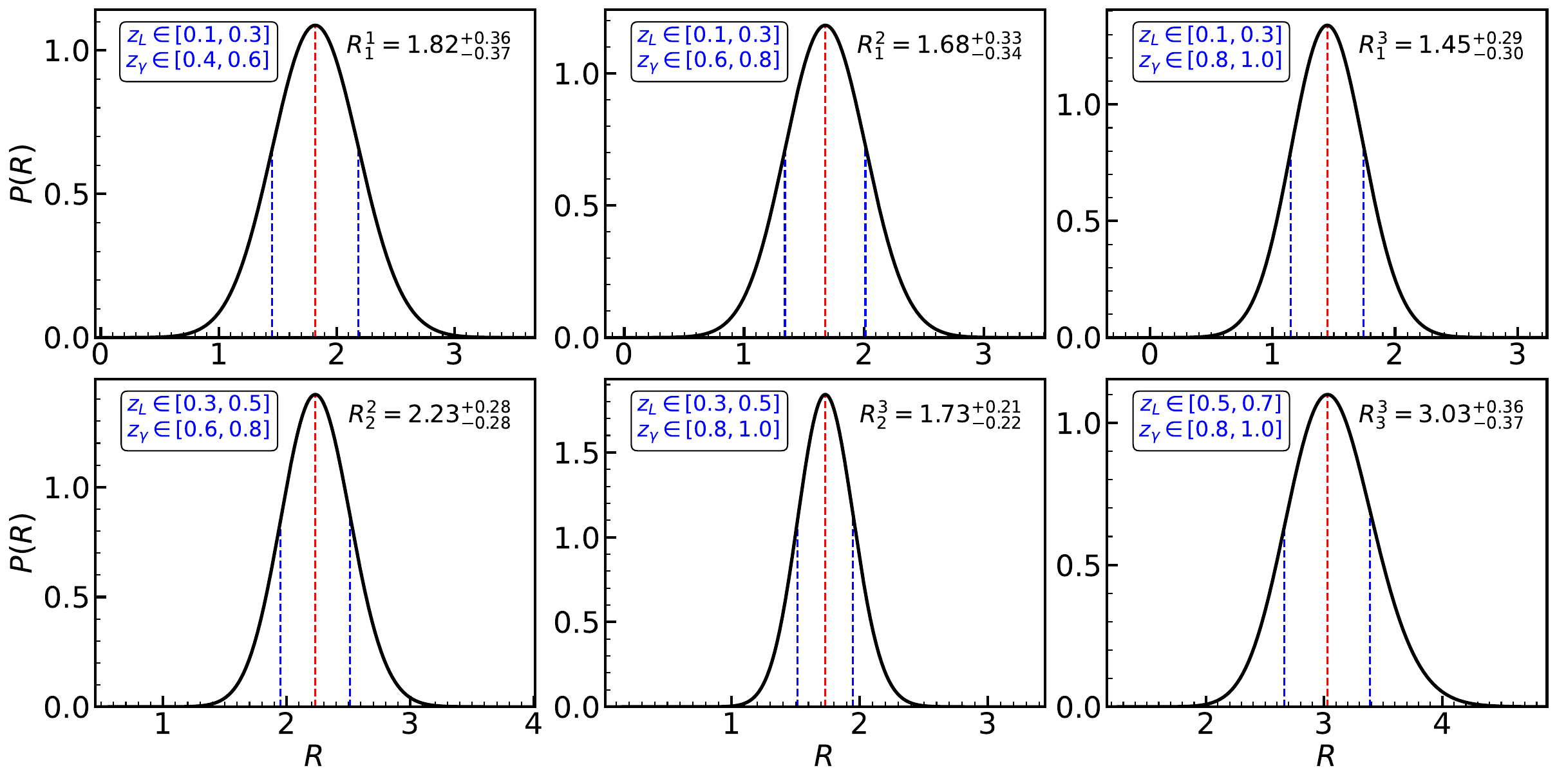}
\caption{Measured posterior PDF $P(R)$ on the lensing ratios, using our ratio measurement method. $P(R)$ is nearly Gaussian. We also show the best-fit value and the associated $1\sigma$ errors.  We take lenses as DESI galaxies at three redshift bins denoted by $z_L$. One of the sources is DECaLS cosmic shear at three redshift bins denoted by $z_\gamma$ and the other is \emph{Planck} CMB lensing. } 
\label{fig:P_R}
\end{figure*}

\subsection{Correlated \texorpdfstring{$\mathbf{d}_{1,2}$}{}}
The above result can be extended straightforwardly to the case of correlated $\mathbf{d}_{1,2}$. Now we define the data vector $\mathbf{d}^T = (\mathbf{d}_1, \mathbf{d}_2)$. The probability distribution of $\mathbf{d}$ is
\begin{equation}
    P(\mathbf{d} | R, \boldsymbol{\lambda}) = \frac{1}{\sqrt{(2\pi)^N {\rm det}\mathbf{C}}} {\rm exp} \left[ -\frac{1}{2} \mathbf{\Delta}^T \mathbf{C}^{-1} \mathbf{\Delta} \right] \ .
\end{equation}
Here $\mathbf{\Delta}\equiv {\bf d}-({\bf A}_1 \boldsymbol{\lambda}, {\bf A}_2\boldsymbol{\lambda})$. 
Because the errors in $\mathbf{d}_{1,2}$ are not independent, the covariance matrix ${\bf C}$ includes the off-diagonal blocks. We denote
\begin{equation}
    \mathbf{C} = \left( 
    \begin{array}{cc}
         \mathbf{C}_{11} & \mathbf{C}_{12} \\
         \mathbf{C}_{21} & \mathbf{C}_{22} 
    \end{array}
    \right) \ ,
    \mathbf{C}^{-1} = \left( 
    \begin{array}{cc}
         \mathbf{B}_{11} & \mathbf{B}_{12} \\
         \mathbf{B}_{21} & \mathbf{B}_{22} 
    \end{array}
    \right) \ ,
\end{equation}
where $\mathbf{C}_{ij} = \langle \mathbf{n}_i \mathbf{n}_j^T \rangle$, and $\mathbf{C}_{12} = \mathbf{C}_{21}$. The blocks of inverse $\mathbf{C}$ are 
\begin{equation}
\begin{aligned}
    &\mathbf{B}_{11} = (\mathbf{C}_{11} - \mathbf{C}_{12} \mathbf{C}_{22}^{-1} \mathbf{C}_{21} )^{-1} \ , \\
    &\mathbf{B}_{12} = -(\mathbf{C}_{11} - \mathbf{C}_{12}\mathbf{C}_{22}^{-1}\mathbf{C}_{21})^{-1} \mathbf{C}_{12} \mathbf{C}_{22}^{-1} \ , \\
    &\mathbf{B}_{21} = -\mathbf{C}_{22}^{-1}\mathbf{C}_{21} (\mathbf{C}_{11} - \mathbf{C}_{12}\mathbf{C}_{22}^{-1}\mathbf{C}_{21})^{-1} \ , \\
    &\mathbf{B}_{22} = \mathbf{C}_{22}^{-1} + \mathbf{C}_{22}^{-1}\mathbf{C}_{21}(\mathbf{C}_{11} - \mathbf{C}_{12} \mathbf{C}_{22}^{-1}\mathbf{C}_{21})^{-1} \mathbf{C}_{12}\mathbf{C}_{22}^{-1} \ .
\end{aligned}
\end{equation}
The expansion of exponential part has 16 terms, which is twice as in Eq.~\ref{eq:expan exp}. We define
\begin{align}
    &\mathbf{Q}' \equiv \mathbf{A}_1^T \mathbf{B}_{11} \mathbf{A}_1 + \mathbf{A}_1^T \mathbf{B}_{12} \mathbf{A}_2 + \mathbf{A}_2^T \mathbf{B}_{21} \mathbf{A}_1 + \mathbf{A}_2^T \mathbf{B}_{22} \mathbf{A}_2 \ , \\
    &\mathbf{T}' \equiv \mathbf{A}_1^T \mathbf{B}_{11} \mathbf{d}_1 + \mathbf{A}_1^T \mathbf{B}_{12} \mathbf{d}_2 + \mathbf{A}_2^T \mathbf{B}_{21} \mathbf{d}_1 + \mathbf{A}_2^T \mathbf{B}_{22} \mathbf{d}_2 \ .
\end{align}
We find that $E$ in Eq. \ref{eqn:PE} is now
\begin{equation}
 E=   -\frac{1}{2} (\boldsymbol{\lambda}^T \mathbf{Q}'\boldsymbol{\lambda} -\boldsymbol{\lambda}^T \mathbf{T}' - \mathbf{T'}^T \boldsymbol{\lambda}) \ . 
\end{equation}
Therefore the final expression of $P(R|{\bf d}_1,{\bf d}_2)$ is identical to Eq. \ref{eq: P_R uncorr}, but replacing ${\bf Q}$ and ${\bf T}$ with ${\bf Q}'$ and ${\bf T}'$. 
\begin{equation}
\label{eq:P_Rcorr}
    P(R|\mathbf{d}) \propto \left[{\rm det}\mathbf{Q}'\right]^{-1/2} {\rm exp} \bigg[ \frac{1}{2} \mathbf{T'}^T \mathbf{Q'}^{-1}\mathbf{T}' \bigg] P_{\rm prior}(R) \ .
\end{equation}

Eq. \ref{eq: P_R uncorr} for uncorrelated ${\bf d}_{1,2}$ and Eq. \ref{eq:P_Rcorr} for correlated ${\bf d}_{1,2}$ are the major results of this paper. 
They provide the analytical expressions of $P(R)$. In order to put the two data sets on an equal footing, one should treat $R$ and $1/R$ equally, which motivates a Jeffreys $1/R$ prior instead of flat prior \footnote{Thanks the anonymous referee for this helpful suggestion. }.

\section{Application: measuring the lensing ratio}
\label{sec: application}
We have applied our method to measure the ratio between cross-correlations of galaxy-ISW and galaxy-CMB lensing \citep{Dong2022}, which is a measure of the gravitational potential decay rate and therefore a measure of dark energy. Here we take the lensing ratio measurement as another example to demonstrate the applicability of our method. 

The two data sets that we adopt are $w^{g\gamma_t}(\theta,z_L,z_\gamma)$ and $w^{g\kappa_{\rm CMB}}(\theta,z_L)$. $w^{g\gamma_t}(\theta,z_L,z_\gamma)$ is the galaxy-tangential shear cross-correlation function between galaxies at lens redshift bin denoted by the mean redshift $z_L$ and shear at source redshift bin denoted by the mean redshift $z_\gamma$. The corresponding galaxy-convergence cross-correlation function is denoted as $w^{g\kappa_\gamma}(\theta,z_L,z_\gamma)$. $w^{g\kappa_{\rm CMB}}(\theta,z_L)$ is the galaxy-CMB lensing convergence cross-correlation function. It is expected that for narrow lens redshift bins \citep{Jain2003, Bernstein2004, ZhangJun2005}, 
\begin{equation}
    w^{g\kappa_{\rm CMB}}(\theta,z_L)=R(z_L,z_\gamma) w^{g\kappa_\gamma}(\theta,z_L,z_\gamma)\ .
\end{equation}
Notice that the lensing ratio $R(z_L,z_\gamma)$ depends on $z_L$ and $z_\gamma$ only through the geometry of the universe, but not structure growth. Measurement errors in $w^{g\gamma_t}(\theta,z_L,z_\gamma)$ and $w^{g\kappa_{\rm CMB}}(\theta,z_L)$ are dominated by shear measurement errors and lensing map noises. So we will treat the two data sets as independent and apply Eq. \ref{eq: P_R uncorr} to measure the ratio $R$. Notice that we restrict the measurement to the CMB lensing/cosmic shear ratio and will not measure the shear ratio between shear at different source redshifts.
In order not to divert readers into  weak lensing details, we present the measurement of $w^{g\gamma_t}(\theta,z_L,z_\gamma)$ and $w^{g\kappa_{\rm CMB}}(\theta,z_L)$ with DESI imaging surveys DR8, DECaLS shear catalog and \emph{Planck} CMB lensing maps in Appendix \ref{sec: data pre-processing}. 

\begin{figure*}
    \centering
    \plotone{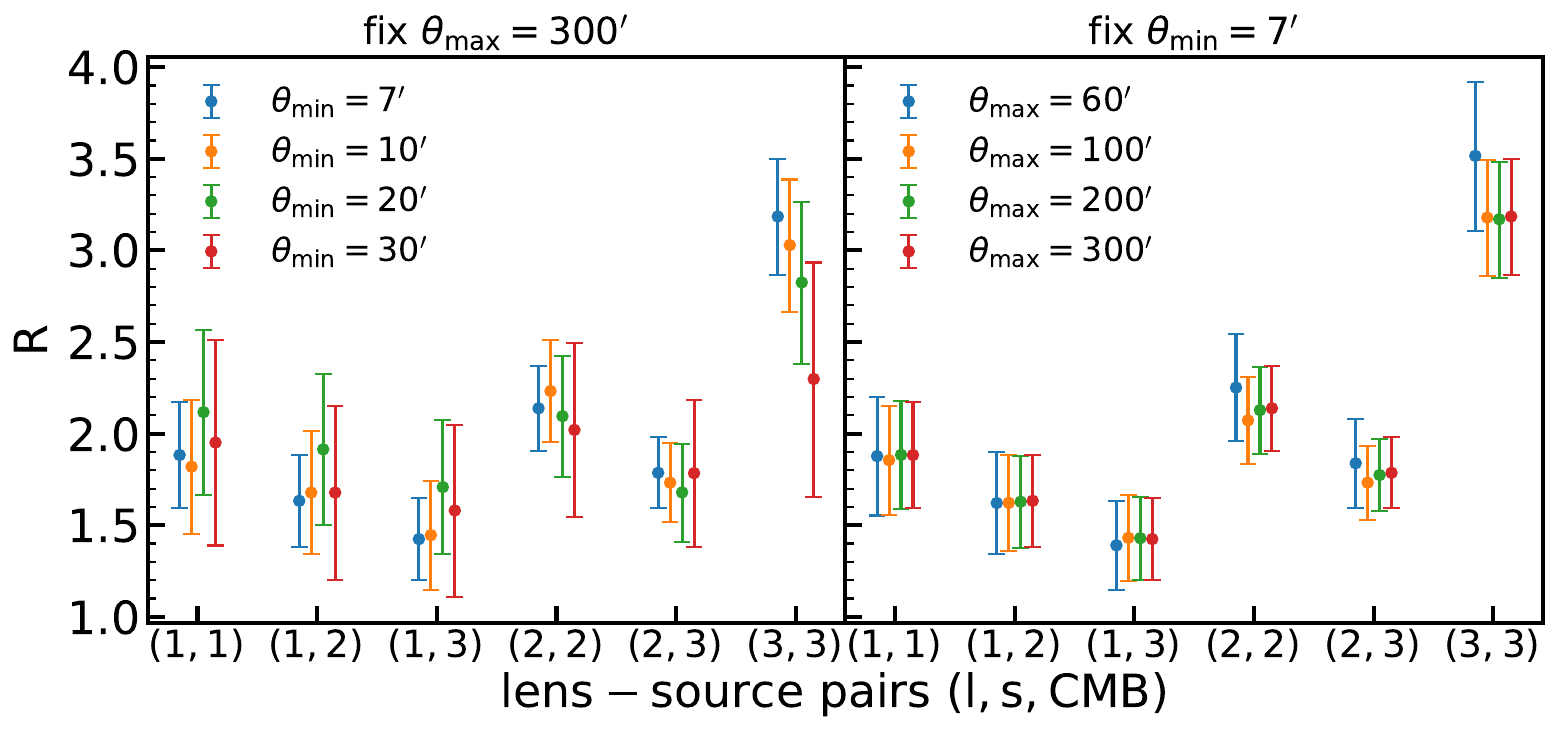}
    \caption{Consistency test 1. We find no statistically significant dependence of $R$ on the chosen scale cut $\theta_{\rm min}$ ($\theta_{\rm max}$).}
    \label{fig:R_theta_cut}
\end{figure*}

Since $w^{g\gamma_t}$ and $w^{g\kappa_{\rm CMB}}$ involve different Bessel functions in the power spectrum-correlation function conversion ($J_2(\ell\theta)$ versus $J_0(\ell\theta)$), they are not proportional to each other. One way to deal with it in our method is to choose the theory $\boldsymbol{\lambda}$ as the power spectrum, and the mapping matrix ${\bf A}_{1,2}$ in Eq. \ref{eq: P_R uncorr} will then involve the oscillating functions of $J_{0,2}$.  This is numerically challenging, even with the help of FFTlog \citep{Hamilton2000}. For the purpose of demonstrating the usage of our method, we adopt a more convenient choice, that is to rescale the correlation functions $w\rightarrow \tilde{w}\equiv w/w_{\rm tem}$. $w_{\rm tem}$ is the corresponding template correlation function based on the theoretical prediction of a fiducial cosmology (Eq.~\ref{eq:w_tem^ggammat} \& Eq.~\ref{eq:w_tem^gkcmb}), which absorbs the $J_{0,2}$ dependences. Therefore, the rescaled correlation functions directly follow the proportionality relation ($\tilde{w}^{g\kappa_{\rm CMB}}=R\tilde{w}^{g\gamma_t}$). Therefore we will use $\tilde{w}^{g\kappa_{\rm CMB}}$ and $\tilde{w}^{g\gamma_t}$ as the data sets to demonstrate our ratio measurement method. 

\subsection{Measurements of lensing ratios}
\label{subsec: measurements of lensing ratios}
Following \S \ref{sec: methodology}, we choose the data sets ${\bf d}_{1,2}$, the theory vector $\boldsymbol{\lambda}$, and the mapping matrixes ${\bf A}_{1,2}$ as 
\begin{equation}
\begin{aligned}
     & \mathbf{d}_1 =\tilde{w}^{g\gamma_t} \ ,  \ \mathbf{d}_2 = \tilde{w}^{g\kappa_{\rm CMB}}\ , \\  & \boldsymbol{\lambda}=\langle {\bf d}_1\rangle\ ,\ {\bf A}_1={\bf I}\ ,\ {\bf A}_2=R{\bf I}\ .
\end{aligned}
\end{equation}
Since the theory vector $\boldsymbol{\lambda}$ is just the expectation value of ${\bf d}_1$, the theory makes no assumptions on cosmology and the measured $R$ will be model independent.  We have three lens bins ($0.1<z_L<0.3$, $0.3<z_L<0.5$ and $0.5<z_L<0.7$) and three source shear bins ($0.4<z_\gamma<0.6$, $0.6<z_\gamma<0.8$ and $0.8<z_\gamma<1.0$). We denote the lens bins with Latin letter $i=1,2,3$ and source bins with Greek letter $\alpha=1,2,3$. Since we only measure the ratio between CMB lensing and cosmic shear, we have six ratios ($R_i^\alpha=R_1^1$, $R_1^2$, $R_1^3$, $R_2^2$, $R_2^3$ and $R_3^3$). We use the data in the range $10 < \theta < 300$ arcmin to measure the ratios.

Fig.~\ref{fig:P_R} shows the posterior $P(R)$, which is normalized such that $\int P(R)dR=1$. $P(R)$ is nearly Gaussian, resulting in $R_1^1 = 1.82^{+0.36}_{-0.37}, R_1^2 = 1.68^{+0.33} _{-0.34}, R_1^3 = 1.45^{+0.29}_{-0.30}, R_2^2 = 2.23^{+0.28}_{-0.28}, R_2^3 = 1.73^{+0.21}_{-0.22}, R_3^3 = 3.03^{+0.36}_{-0.37}$. The S/N of each ratio is between $4.9$ and $8.4$.  The error budget is dominated by errors in the galaxy-CMB lensing correlation measurement. Therefore errors in $R_1^1$, $R_1^2$ and $R_1^3$ are tightly correlated, since they share the same galaxy-CMB lensing measurement. So are errors in $R_2^2$ and $R_2^3$.  Meanwhile, we test our method for multiple correlated ${\bf R_1^\alpha}={\rm diag}(R_1^1,...,R_1^2,...,R_1^3,...)$ and ${\bf R_2^\alpha}={\rm diag}(R_2^2,...,R_2^3,...)$. We take the vector of $\tilde{w}^{g\gamma_t}$ in the identical lens bin as ${\bf d}_{1}$, and the vector of $\tilde{w}^{g\kappa_{\rm CMB}}$ in the identical lens bin as ${\bf d}_{2}$. For example, ${\bf d}_{1} = (\tilde{w}^{g\gamma_t,1}_1, \tilde{w}^{g\gamma_t,2}_1, \tilde{w}^{g\gamma_t,3}_1)$, ${\bf d}_{2} = (\tilde{w}^{g\kappa_{\rm CMB}}_1, \tilde{w}^{g\kappa_{\rm CMB}}_1, \tilde{w}^{g\kappa_{\rm CMB}}_1$). The theory vector $\boldsymbol{\lambda}=\langle {\bf d}_1\rangle$, and the mapping matrices ${\bf A}_1={\bf I}$, $ {\bf A}_2={\bf R_1^\alpha I}$. The measured ratios are identical with those shown in Fig.~\ref{fig:P_R}.

\subsection{Consistency checks}
\label{subsec: consistency checks}

To demonstrate the validity of our method and measurement, we perform several consistency checks.

We first check whether the constrained $R$ depends on the chosen $\theta$ range ($\theta_{\rm min} < \theta < \theta_{\rm max}$). By theoretical design, the ratio $R$ is scale-independent. Therefore the measured $R$ should be independent of scale cuts.  Nonetheless, the correlation function measurements themselves may suffer from potential systematics in the galaxy clustering, shear, and CMB lensing. The left panel of Fig.~\ref{fig:R_theta_cut} shows the consistency tests by varying $\theta_{\rm min} = 7', 10', 20', 30'$ while fixing $\theta_{\rm max} = 300'$. The right panel shows the tests by varying $\theta_{\rm max} = 60', 100', 200', 300'$ while fixing $\theta_{\rm min} = 7'$. The constrained $R$ are fully consistent with each other. The left panels show a larger scatter in $R$, since a small scale cut affects the overall S/N more significantly. 

\begin{figure}[!htb]
    \centering
    \includegraphics[width=\columnwidth]{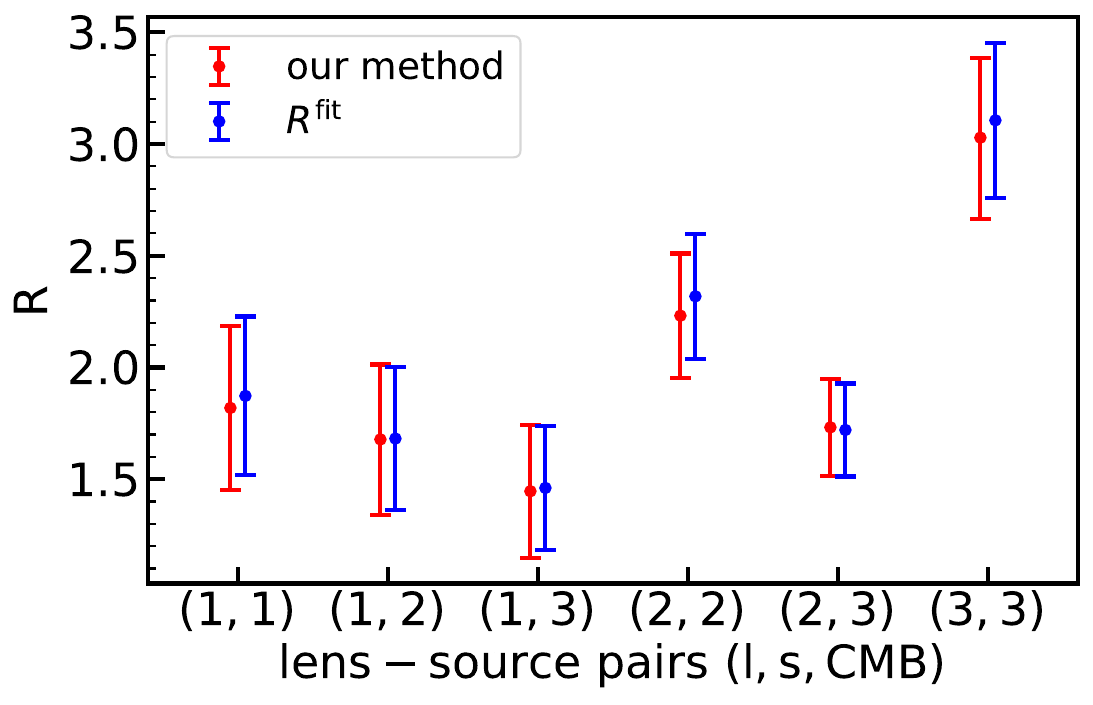}
    \caption{Consistency test 2. We compare the measured $R$ by our method with $R^{\rm fit}$ from model fitting.  }
    \label{fig:R_fit}
\end{figure}

The second check is to compare against the direct model fitting result introduced in \S \ref{subsec: sn}. This method adopts the \emph{Planck} cosmology predicted $w$ up to a scale-independent amplitude $b$ and fit $b$ for both $w^{g\gamma_t}$ and $w^{g\kappa_{\rm CMB}}$.  The ratio is then  $R^{\rm fit} = b_{\rm fit}^{g\kappa_{\rm CMB}}/b_{\rm fit}^{g\gamma_t}$. The error of $R^{\rm fit}$ is 
\begin{equation}
\Delta R^{\mathrm{fit}}=\frac{b_{\mathrm{fit}}^{g \kappa_{\mathrm{CMB}}}}{b_{\mathrm{fit}}^{g \gamma_t}} \sqrt{\left(\frac{\sigma_b^{g \kappa_{\mathrm{CMB}}}}{b_{\mathrm{fit}}^{g \kappa_{\mathrm{CMB}}}}\right)^2+\left(\frac{\sigma_b^{g \gamma_t}}{b_{\mathrm{fit}}^{g \gamma_t}}\right)^2} \ .
\end{equation} 
The values of $R^{\rm fit}$ are $R_1^{1, \rm fit} = 1.87 \pm 0.36, R_1^{2, \rm fit} = 1.68 \pm 0.32, R_1^{3, \rm fit} = 1.46 \pm 0.28, R_2^{2, \rm fit} = 2.32 \pm 0.28, R_2^{3, \rm fit} = 1.72 \pm 0.21, R_3^{3, \rm fit} = 3.11 \pm 0.35$. 
As shown in Appendix \ref{subsec: sn}, this one-parameter bias model describes the measured correlation functions well, and confirms the robustness of the cross-correlation measurement. Therefore the ratio obtained in this way provides a robust check of $R$ measured by our method. Fig.~\ref{fig:R_fit} shows the two results (best-fit values and the associated errors) are fully consistent with each other. Our method, without the assumption of scale-independent bias, can be directly applied to smaller angular scales, where scale dependence of bias may be non-negligible. 

\begin{figure}[!htb]
    \centering
    \includegraphics[width=\columnwidth]{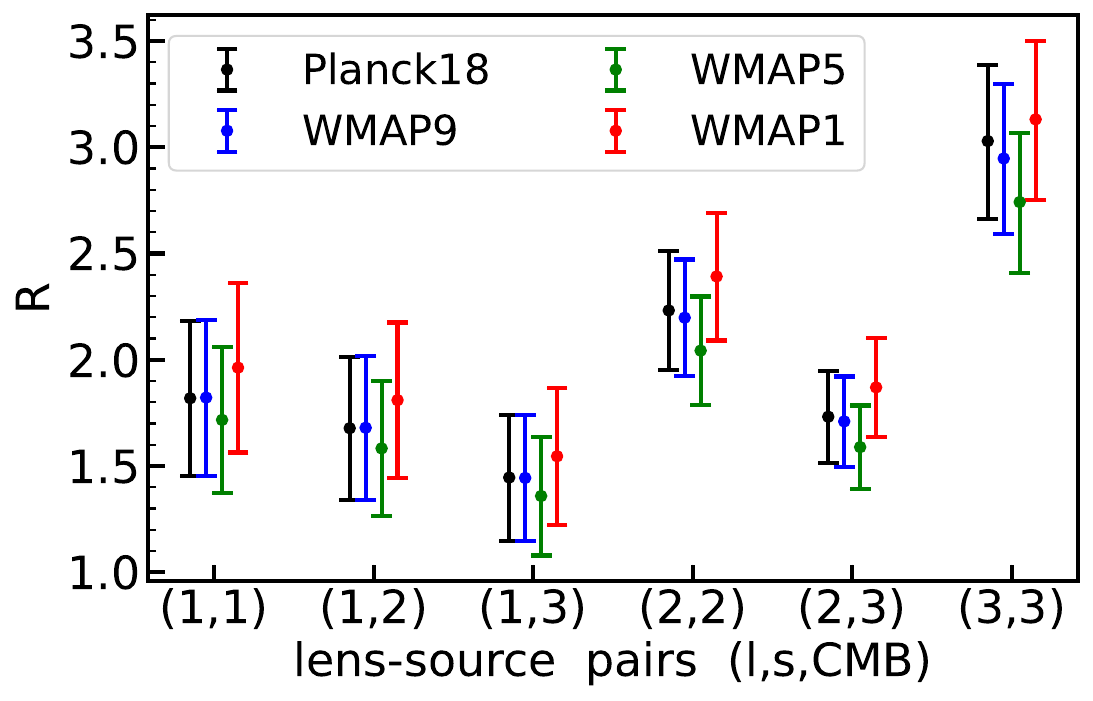}
    \caption{Consistency test 3. The measured $R$ has negligible dependence on the adopted theoretical template, which depends on cosmology.  }
    \label{fig:R_wmap}
\end{figure}

The third test is on the potential cosmological dependence that we introduce by the scaling $w\rightarrow w/w_{\rm tem}$, which is a convenient but not exact implementation of our method. We measure the lensing ratio adopting different cosmologies (the first year \citep{wmap1}, five-year \citep{wmap5},  nine-year \emph{WMAP} cosmology \citep{wmap9} and \emph{Planck} 2018 cosmology \citep{Planck2018parameters} in Table.~\ref{tab:parameter}) and corresponding template of scaling. Fig.~\ref{fig:R_wmap} shows that, despite various differences in cosmological parameters, the differences in the measured $R$ are all smaller than $0.5\sigma$ and therefore are totally negligible. 

\begin{table}[!htb]
    \centering
    \begin{tabular}{cccccccc} 
    	\hline
    	Parameter &  $\Omega_{\rm c}$ & $\Omega_{\rm b}$ & $n_s$ & $H_0$ & $\sigma_8$ \\
    	\hline 
        WMAP1 & 0.224 & 0.0463 & 0.99 & 72 & 0.9  \\
        WMAP5 & 0.206 & 0.0432 & 0.961 & 72.4 & 0.787 \\
        WMAP9 & 0.235 & 0.0464  & 0.9710 & 69.7 & 0.820 \\
        \emph{Planck}18 & 0.265 & 0.04887 & 0.9649 & 67.36 & 0.8111  \\
    	\hline
    \end{tabular}
    \caption{The first year \citep{wmap1}, five-year \citep{wmap5}, nine-year \emph{WMAP} cosmology \citep{wmap9}, and \emph{Planck} 2018 cosmology \citep{Planck2018parameters}. }
    \label{tab:parameter}
\end{table}

\section{Conclusions} \label{sec: conclusions}
We develop an unbiased method to solve the problem of measuring the ratio of two data sets. This solution is developed based on Bayesian analysis, including all the data points and their uncertainties. The posterior distribution of the ratio $P(R)$ has an analytical expression. This method enables fast and unbiased $R$ measurement, with minimal statistical errors. Furthermore, it relies on the usual assumption of Gaussian error in data, but no underlying model other than the proportionality relation between the two data sets.

We measure the lensing ratio as an application. We take the lenses as DESI imaging survey galaxies, and sources as DECaLS cosmic shear and \emph{Planck} CMB lensing. We measure the ratio between CMB lensing and cosmic shear at multiple lens-source redshift pairs, with S/N ranging from 5 to 8. We verify that the measured $R$ is insensitive to the scale cuts and the adopted cosmology. Together with another example of measuring the decay rate of cosmological gravitational potential \citep{Dong2022}, we demonstrate the applicability of our method to measure the ratios.

\section*{Acknowledgements}
This work is supported by the National Science Foundation of China (11621303), the National Key R\&D Program of China (2020YFC2201602, 2018YFA0404504, 2018YFA0404601, 2020YFC2201600) and CMS-CSST-2021-A02. J.Y. acknowledges the support from China Postdoctoral Science Foundation (2021T140451). H.Y.S. acknowledges the support from CMS-CSST-2021-A01, NSFC of China under grant 11973070, and Key Research Program of Frontier Sciences, CAS, grant No. ZDBS-LY-7013. This work made use of the Gravity Supercomputer at the Department of Astronomy, Shanghai Jiao Tong University. The results in this paper have been derived using the following packages: Numpy \citep{numpy}, \texttt{HEALPix} \citep{healpix}, IPython \citep{ipython}, CCL \citep{pyccl}, and TreeCorr \citep{treecorr}.


\section*{Data Availability}
The Python code, \texttt{Jupyter} notebooks, and the data files for reproducing the results and figures of this work can be found on \texttt{Github} at \url{https://github.com/Ze-yangSUN/Ratio_method}.

\appendix

\section{Data Pre-processing} 
\label{sec: data pre-processing}
We introduce the data that we use to measure the cross-correlation functions of galaxy-tangential shear, and galaxy-CMB lensing. These cross-correlations will then be used by our ratio estimator as the input. 

\subsection{The lens galaxy catalog}
\label{subsec: lens galaxy catalog}

For the lens galaxies, we choose DESI imaging survey DR8 \citep[Dark Energy Camera Legacy Survey;][]{Dey2019}. We also use the photometric redshift measurement provided by \cite{Zou2019}. This catalog also applies to other cosmological analyses (e.g. \citet{Yao2020SC, Zhangziwen2021, Dong2021LDP, Zou2021}). The imaging footprints cover an area of over 14,000 deg$^2$ in both northern and southern Galactic caps. The catalog contains about 0.17 billion morphologically classified galaxies with $r < 23$ mag and covers the redshift range of $z^P < 1$.

We select the lens galaxies in three redshift bins of $0.1<z_L<0.3$, $0.3<z_L<0.5$ and $0.5<z_L<0.7$, with $r$-band apparent magnitude cut of 18.5, 19.5, 20.5 at each bin respectively. The final catalog contains $\sim 3.64, 2.59, 2.20$ million galaxies in each bin and $\sim 8.43$ million in total. To select the random galaxies\footnote{ \url{https://www.legacysurvey.org/dr8/files/\#random-catalogs}} with considering the survey geometry, we first generate a mask of lens galaxies, with \texttt{nside} = 512 in \texttt{HEALPix} \citep{healpix}. The binary tracer mask is 1 when there are tracer galaxies located at the pixel; otherwise it is 0. The random galaxies have an identical mask to the tracer galaxies. 

\subsection{The weak lensing catalogs}
\label{subsec: weak lensing catalogs}
For weak lensing, we use two data sets. One is the shear catalog derived from the DECaLS images of DR8.  The shear catalog models potential biases with a multiplicative and an additive bias \citep{Heymans2012, Miller2013}. The shear measurement and imperfect modeling of point-spread function (PSF) size result in the multiplicative bias. To calibrate our shear catalog, we cross-matched the DECaLS DR8 objects with the external shear measurements, including Canada-France-Hawaii Telescope (CFHT) Stripe 82, Dark Energy Survey  \citep[DES;][]{DES2016} and Kilo-Degree Survey \citep[KiDS;][]{Hildebrandt2017}. The additive bias comes from residuals in the anisotropic PSF correction which depends on galaxy sizes. The additive bias is subtracted from each galaxy in the catalog. The DECaLS DR8 shear catalog was used in the cluster weak lensing measurement by \cite{Zu2021}, and the same shear catalog and photo-$z$ measurements (but from DECaLS DR3) were used in the weak lensing analysis of intrinsic alignment studies of \cite{Yao2020SC}, and CODEX clusters by \cite{Phriksee2020}, and we refer readers to these two papers for technical details of the DECaLS shear catalog and photo-z errors. The shear catalog contains $\sim 43.9$ million galaxies covering $\sim$ 15,000 deg$^2$. We split it into three redshift bins of $0.4<z_\gamma<0.6$, $0.6<z_\gamma<0.8$, $0.8<z_\gamma<1.0$. These bins have $26.9$, $11.8$, $5.18$ million source galaxies respectively.  

\begin{figure*}
    \centering
    \includegraphics[width=2\columnwidth]{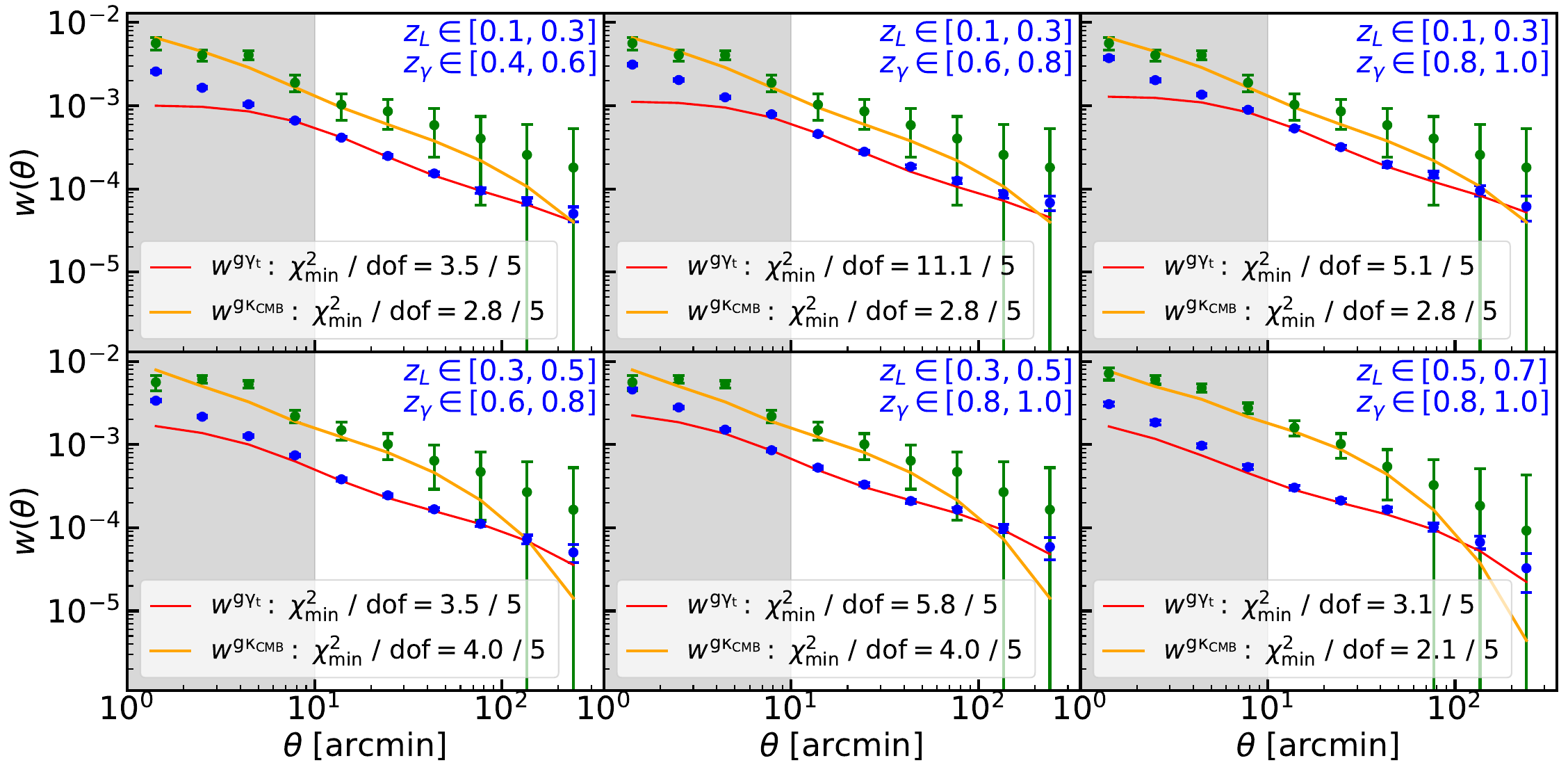}
    \caption{Measurements of galaxy-shear cross-correlation ($W^{g\gamma_t}$) and galaxy-CMB lensing cross-correlation ($w^{g\kappa_{\rm CMB}}$) for various lens-source pairs, together with the best-fitting theory curves. The gray areas ($\theta < 10'$) correspond to the scales that are not used in the lensing ratio measurement.} 
    \label{fig:corr}
\end{figure*}
Another lensing data is the latest \emph{Planck} CMB lensing convergence maps with corresponding mask \citep{planck2020b, Planck2018parameters}. As we have found in our previous work of DESI galaxy group-Planck CMB lensing cross-correlation \citep{Sun2022}, the residual thermal Sunyaev-Zel'dovich (SZ) effect contaminates the lensing map and biases the cross-correlation measurement. So in the current analysis we adopt the CMB lensing map reconstructed from the tSZ-deprojected temperature-only SMICA map\footnote{\url{http://pla.esac.esa.int/pla/\#cosmology}} \citep{planck2020b}. The spherical harmonic coefficients of lensing convergence are provided in \texttt{HEALPix} with $\ell_{\rm max} = 4096$, and the associated mask is provided with the resolution \texttt{nside} = 2048. Due to the overwhelming reconstruction noise at small scales, the $\kappa$ map was filtered to remove modes with $\ell > 1536$. Since we focus on large-scale clustering, we downgraded the $\kappa$ map to \texttt{nside} = 512 with $\ell_{\rm max}$ = 1536.

\subsection{Measurements of the correlation functions} \label{subsec: two-point}
We measure the angular two-point correlation function between the pixelized CMB lensing convergence map $\kappa_{\rm CMB}$ and the galaxy overdensity $\delta_g$ by summing over tracer-convergence pixel pairs $g$, separated by angle $\theta$. We subtract the corresponding correlation with a sample of random points in place of the tracer galaxies, where the sum is over random-convergence pairs $r$ separated by $\theta$. The final estimator is
\begin{equation}
    w^{g\kappa_{\rm CMB}}(\theta) = \frac{\sum_g \omega_g\kappa_g}{\sum_g \omega_g}(\theta) - \frac{\sum_r \omega_r \kappa_r}{\sum_r \omega_r}(\theta) \ ,
\end{equation}
where $\omega_g$ and $\omega_r$ are the weights associated respectively with each tracer galaxy and random point. For the random points we set $\omega_r = 1$, and for the galaxies we set $\omega_g = 1$ without considering the observational systematics. This estimator is analogous to that used in tangential shear measurements in galaxy-galaxy lensing, which is 
\begin{equation}\label{eq:wggammat obs}
    w^{g\gamma_t} = \frac{\sum_{\rm ED} w_j \gamma_j^+}{\sum_{\rm ER} (1+m_j)w_j}  \ ,
\end{equation}
where $\sum_{\rm ED}$ means summing over all the tangential ellipticity (E)-galaxy count in the data (D) pairs, $\sum_{\rm ER}$ means summing over all the tangential ellipticity (E)-galaxy count in the random catalog (R) pairs. The random catalog aims to subtract the selection bias in the correlation function from the shape of the footprint of the survey. Besides, $w_j$ denotes the \textit{lens}fit \citep{Miller2013} weight for the galaxy shape measurement; $\gamma_j^+$ is the tangential ellipticity for the $j$th galaxy; $m_j$ is the estimation of  multiplicative bias provided by the same catalog. 

For the fiducial measurements in this work, we group the tracer-convergence / tracer-shear pairs in 10 log-spaced angular separation bins between $1'$ and $300'$. We use \texttt{treecorr} \citep{treecorr} to measure all two-point correlation functions and covariance matrices. We estimate the covariance matrix between the measurements using the jackknife method. In this approach, the survey area is divided into 100 regions (`jackknife patches'), and the correlation function measurements are repeated once with each jackknife patch removed for the tracer as well as the lensing sample. The measured correlation functions between galaxy-galaxy lensing or galaxy-CMB lensing are shown as a function of angular scale in Fig.~\ref{fig:corr}.

\subsection{Evaluating the S/N of correlation measurements}
\label{subsec: sn}
We first quantify the detection significance of a non-zero signal, $(S/N)^2\equiv \chi^2_{\rm null}$. Here 
\begin{equation}\label{eq:chi2_null}
    \chi^2_{\rm null} = \sum\limits_{\theta\theta^\prime} w_{\rm obs}(\theta) \textbf{Cov}_{\theta\theta^{\prime}}^{-1} w_{\rm obs}(\theta^\prime) \ . 
\end{equation}
$w_{\rm obs}(\theta)$ is for observation data, and $\textbf{Cov}_{\theta\theta^{\prime}}$ is the covariance matrix via the jackknife method. The detections are significant at all six lens-source pairs, $S/N = 30.7, 30.5, 28.2, 31.5, 31.0, 19.3$ for 11, 12, 13, 22, 23, 33 galaxy-shear pairs, their total $S/N$ is 70.7. $S/N = 5.6, 8.8, 10.2$ for galaxy-$\kappa_{\rm CMB}$ from low to high redshift, respectively, their total $S/N$ is 14.6. To evaluate the above $S/N$, we use the data in the range $10' < \theta < 300'$. Therefore the error budget of lensing ratio measurement is dominated by that in galaxy-CMB lensing, which is in turn dominated by noises in the CMB lensing map.

\begin{figure*}
    \centering
    \includegraphics[width=2\columnwidth]{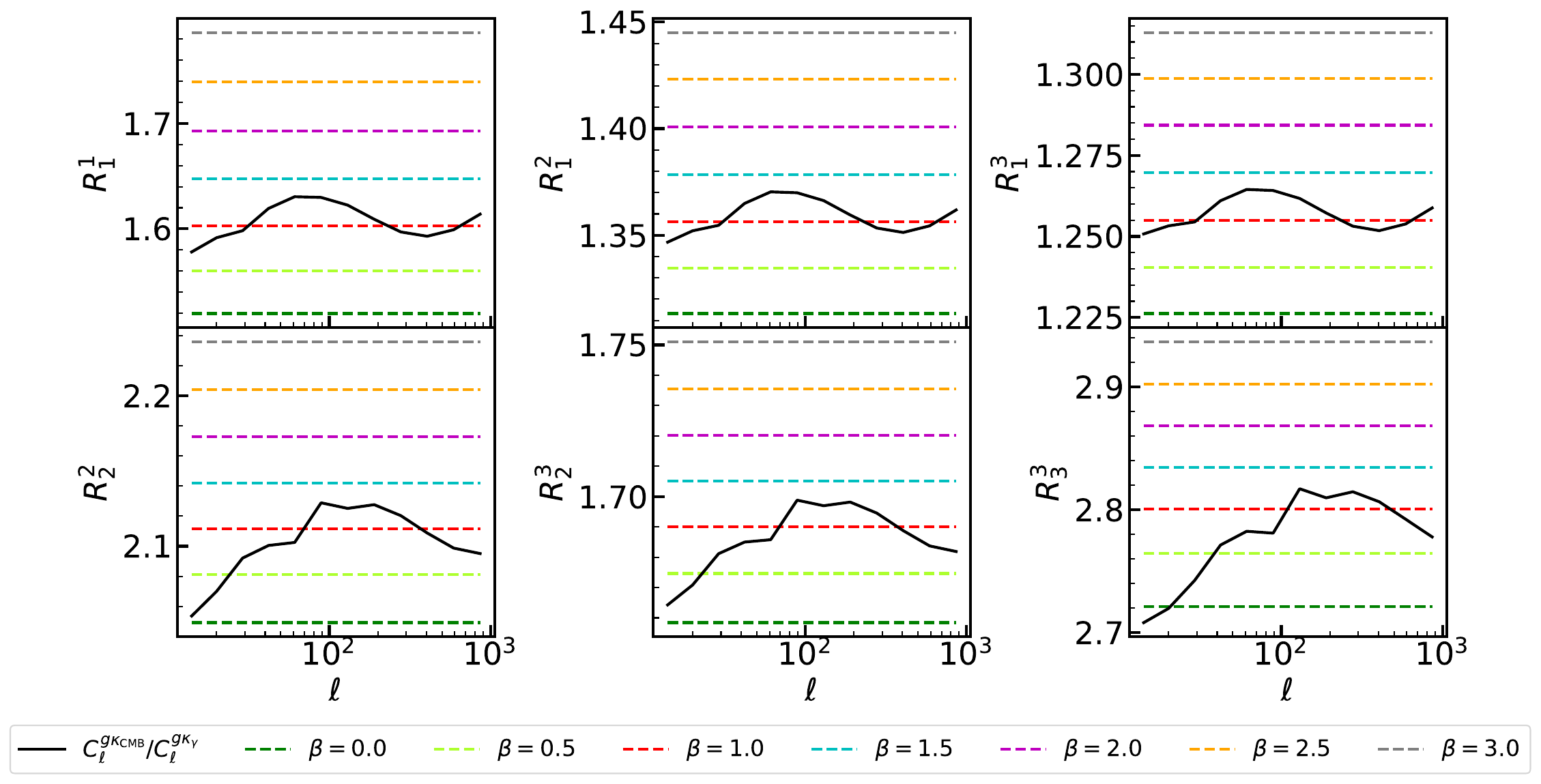}
    \caption{The exact lensing ratio ( $C_\ell^{g\kappa_{\rm CMB}} / C_\ell^{g\kappa_\gamma}$) and its modeling (Eq.~\ref{eq:modeling the R} with different values of $\beta$). Eq.~\ref{eq:modeling the R} with $\beta = 1.0$ provides an excellent approximation of the lensing ratio. }
    \label{fig:R_beta}
\end{figure*}

We further check whether the measurements agree with our theoretical expectation, by the goodness of fitting. We approximate the galaxy bias defined through $b_g(k)\equiv P_{gm}(k)/P_m(k)$ as scale-independent. Here $P_m$ is the nonlinear matter power spectrum. Under this approximation,  $C^{g\kappa}=bC^{m\kappa}$. We fix the cosmology as \emph{Planck} 2018 cosmology \citep{Planck2018parameters}: $\Omega_m = 0.315, \Omega_\Lambda = 0.685, n_s = 0.965, h = H_0/(100 \ \rm km \ s^{-1} \ Mpc^{-1}) = 0.674$ and $\sigma_8 = 0.811$. We then  compute $C^{m\kappa}$ using the Core Cosmology Library (CCL; \citet{pyccl}). We then convert $C^{g\kappa}$ into $w^{g\gamma_t}$ and $w^{g\kappa_{\rm CMB}}$ with FFTLog in \texttt{pyccl} \citep{pyccl}. Since in the CMB lensing correlation measurement we have filtered away modes $\ell \ga  1536$ to suppress measurement noise, we adopt the same maximum $\ell_{\rm max}=1536$ cut in the integral of Eq. \ref{eq:wgkcmb}. We then have the theoretical template for the one-parameter fitting, $w=bw_{\rm tem}$ (Eq.~\ref{eq:w_tem^ggammat} \& Eq.~\ref{eq:w_tem^gkcmb}).  We adopt the same power spectrum for generating $w^{g\gamma_t}_{\rm tem}$ and $w^{g\kappa}_{\rm tem}$. So Fig.~\ref{fig:corr} also shows the best-fit theoretical curves. Since the constant bias approximation only holds at relatively large scale, we take a scale cut  $10' < \theta < 300'$, resulting in satisfying  $\chi^2_{\rm min}$/d.o.f.$\la 1$ (Fig.~\ref{fig:corr}). We are then confident that at least within this scale range, the measurements have insignificant contamination.

\section{The lensing ratio statistics} 
\label{sec: statistics}
The lensing convergence $\kappa$, in the direction $\boldsymbol{\theta}$ is given by
\begin{equation}
    \kappa(\boldsymbol{\theta}) = \int_0^{\infty}d\chi W(\chi)\delta(\boldsymbol{\theta}, \chi) \ .
\end{equation}
Here $\delta(\chi,\boldsymbol{\theta})$ is the density fluctuations at sky position $\boldsymbol{\theta}$.  $W(\chi)$ is the $\kappa$ kernel. The one corresponding to the galaxy shear catalog is 
\begin{equation}\label{eq:Wk}
    W^\gamma(\chi) = \frac{3}{2}\Omega_{m0} H_0^2 (1+z) \int_\chi^\infty d\chi' n_\gamma(\chi') \frac{d_A(\chi) d_A(\chi, \chi')}{d_A(\chi')} \ .
\end{equation}
Here $\Omega_{m0}$ is the matter density today, $H_0$ is the Hubble constant today, $\chi$ is the comoving distance to redshift $z$, $d_A(\chi)$ is the angular diameter distance to $\chi$, $d_A(\chi,\chi')$ is the angular diameter distance between comoving distance $\chi$ and $\chi'$, and $n_\gamma(\chi)$ represents the normalized distribution of source shear galaxies.

For CMB lensing, the source distribution can be approximated as a Dirac $\delta_D$ function centered at the comoving distance to the surface of last scattering, $\chi^* = \chi(z^*\approx1090)$. In this case, the lensing kernel $W$ is given by
\begin{equation}\label{eq:Wkcmb}
    W^{{\rm CMB}}(\chi) = \frac{3}{2}\Omega_{m0}H_0^2(1+z)\frac{d_A(\chi) d_A(\chi, \chi^*)}{d_A(\chi^*)}  \ .
\end{equation}

Using the Limber approximation \citep{Limber1953} and the flat sky approximation, the galaxy overdensity-lensing cross power spectrum is
\begin{equation}
    C^{g\kappa}(\ell) = \frac{2\pi^2}{\ell^3}\int d\chi n_L(z) W(\chi) d_A(\chi) \Delta_{\rm gm}^2\left(k = \frac{\ell+1/2}{d_A(\chi)},z\right) \ .
\label{eqn:ckg}
\end{equation}
Here $\Delta_{\rm gm}^2(k) = k^3P_{\rm gm}(k)/(2\pi^2)$ and $P_{\rm gm}$ is the 3D galaxy-matter cross power spectrum.  $n_L(z)$ is the redshift distribution for the lens galaxies. Notice that $W(\chi)$ can either be $W^\gamma(\chi)$ or $W^{\rm CMB}$. The corresponding galaxy-tangential shear cross-correlation and galaxy-CMB lensing convergence cross-correlation functions are
\begin{equation}
\label{eq:wggammat}
    w^{g\gamma_t}(\theta) = \int_0^\infty \frac{\ell d\ell}{2\pi} J_2(\ell\theta) C^{g\kappa_\gamma}(\ell) \ ,
\end{equation}
\begin{equation}
\label{eq:wgkcmb}
    w^{g\kappa_{\rm CMB}}(\theta) = \int_0^{\infty} \frac{\ell d\ell}{2\pi} J_0(\ell\theta) C^{g\kappa_{\rm CMB}}(\ell) \ .
\end{equation}
Here  $J_0$ and $J_2$ are the zero- and second-order Bessel functions of the first kind, respectively.  $J_2$, instead of $J_0$, shows up in Eq.~\ref{eq:wggammat}, since shear is a spin-2 field and only the tangential shear is correlated with the galaxy distribution (scalar field).  

The corresponding template correlation functions $w_{\rm tem}$ of above two cross-correlations are
\begin{equation}
\label{eq:w_tem^ggammat}
    w_{\rm tem}^{g\gamma_t} (\theta) = \int_0^{\ell_{\rm max}} \frac{\ell d\ell}{2\pi} J_2(\ell\theta) C^{g\kappa_{\rm CMB}}_{\rm tem}(\ell) \ ,
\end{equation}

\begin{equation}
\label{eq:w_tem^gkcmb}
    w_{\rm tem}^{g\kappa_{\rm CMB}} (\theta) = \int_0^{\ell_{\rm max}} \frac{\ell d\ell}{2\pi} J_0(\ell\theta) C^{g\kappa_{\rm CMB}}_{\rm tem}(\ell) \ .
\end{equation}
It should be noted that $w_{\rm tem}^{g\gamma_t}$ is not converted from its corresponding power spectrum $C_{\rm tem}^{g\kappa_\gamma}$ in Eq.~\ref{eq:w_tem^ggammat}. Because we adopt the same power spectrum $C_{\rm tem}^{g\kappa_{\rm CMB}}$ for generating $w_{\rm tem}^{g\gamma_t}$ and $w_{\rm tem}^{g\kappa_{\rm CMB}}$. This causes the rescaled correlation functions $\tilde{w} (\equiv w/w_{\rm tem})$ to follow the proportionality relation $\tilde{w}^{g\kappa_{\rm CMB}}=R\tilde{w}^{g\gamma_t}$. 

\subsection{Modeling the lensing ratio}
\label{subsec: modeling}
For the same lens sample of sufficiently narrow redshift distribution, we expect 
\begin{equation}
\label{eqn:R}
    R=\frac{C^{g\kappa_{\rm CMB}}}{C^{g\kappa_\gamma}}=\frac{W^{\rm CMB}(\bar{\chi}_L)}{W^\gamma(\bar{\chi}_L)}\ .
\end{equation}
Here $\bar{\chi}_L$ is the mean radial coordinate of lens galaxies. The ratio does not depend on the matter clustering and complexities associated with it, so it provides a clean measure of cosmic distances. The key here is the cancellation of the galaxy-matter power spectra in Eq. \ref{eqn:ckg} \& Eq. \ref{eqn:R}. This requires that the lens galaxies are narrowly distributed in redshift space. Under such a limit, the power spectrum can be well approximated as a constant and can be moved outside of the integral. This approximation breaks to a certain extent in our analysis, since the photometric redshift width is $\Delta z^P_L=0.2$ and the true redshift width is expected to be larger. Scrutinizing Eq. \ref{eqn:ckg} \& Eq. \ref{eqn:R}, we expect
\begin{equation}\label{eq:modeling the R} 
    R = \frac{\int_0^\infty (1+z) \frac{d_L d_{L\rm CMB}}{d_{\rm CMB}} n_L(z)\chi^{-2+\beta} dz}{\int_0^\infty (1+z) d_L n_L(z) \chi^{-2+\beta} dz  \frac{\int_z^\infty\frac{d_{L\gamma}}{d_\gamma}n_\gamma(z_\gamma)dz_\gamma}{\int_0^\infty n_\gamma(z_\gamma) dz_\gamma}} \ .
\end{equation}
Here the parameter $\beta$ and the associated function $\chi^{-2+\beta}$ are to weigh the contribution from each redshift. $\beta=0$ corresponds to approximate $P_{\rm gm}(k=\ell/d_A(z),z)$ in the integrals of Eq. \ref{eqn:ckg} \& Eq. \ref{eqn:R} as redshift independent, for fixed $\ell$. $\beta=3$ corresponds to approximate  $\Delta^2_{\rm gm}(k=\ell/d_A(z),z)$ as redshift independent. There is no guarantee on either choice of $\beta$. So we try  various values of parameter $\beta\in [0,3]$  and compare with $R$ calculated as $C^{g\kappa_{\rm CMB}}/C^{g\kappa_\gamma}$.  Fig.~\ref{fig:R_beta} shows that the choice of $\beta=1$ is nearly exact. This choice of $\beta=1$ is insensitive to the redshift pairs, as shown in the same figure. Furthermore, we have tested that even if we enlarge the adopted photo-z error by a factor of $2$, $\beta=1$ remains the most accurate.  Therefore, we adopt $\beta = 1.0$ throughout this work to calculate the theoretical expectation value of ratios.

\subsection{Implications of the ratio measurement}
\label{subsec: implications}

\begin{figure}
    \centering 
    \includegraphics[width=\columnwidth]{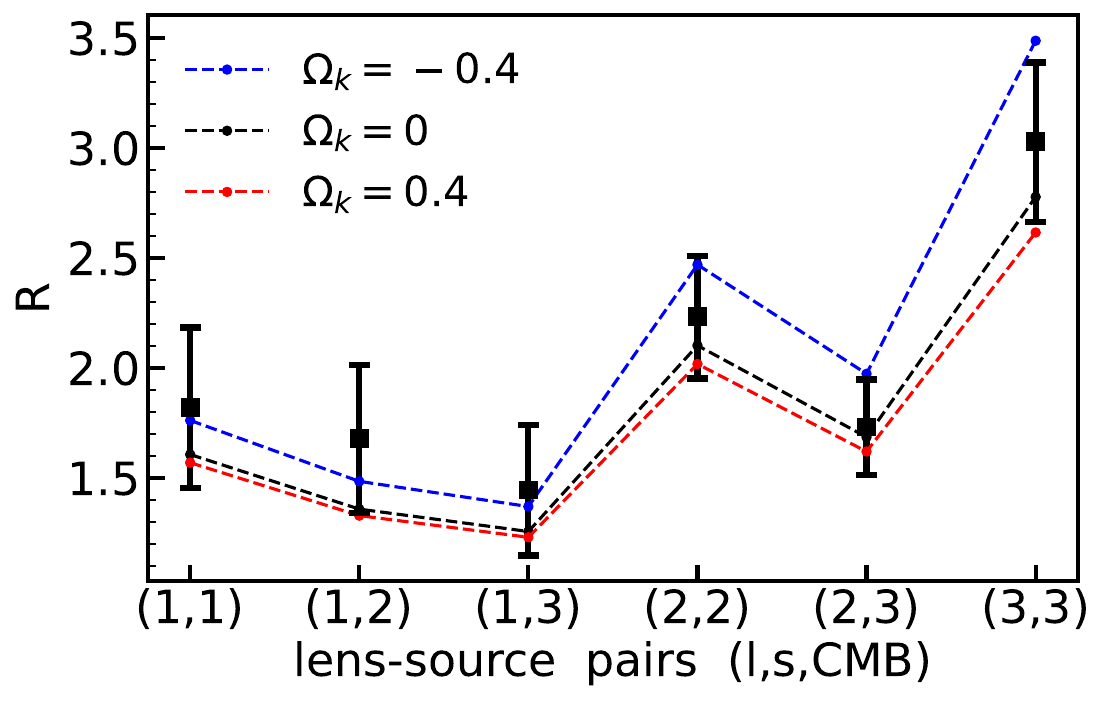}
    \caption{The impact of curvature $\Omega_k$ on the lensing ratios $R$. The negative value of $\Omega_k$ has a significantly greater impact on the lensing ratios than the positive value. }
    \label{fig:R_Omegak}
\end{figure}

We now proceed to the comparison between the measured $R$ and the theoretical prediction of the fiducial flat \emph{Planck} 2018 cosmology. Fig.~\ref{fig:R_Omegak} shows that the two agree with each within the measurement statistical error bars.  Nevertheless, the theory seems systematically lower than the data. The difference is $17\%$ in $R_1^1$, $R_1^2$ \& $R_1^3$, $4\%$ in $R_2^2$ \& $R_2^3$, and $8\%$ in $R_3^3$. Although the significance of the overall discrepancy is only $\Delta \chi^2=1.04$.\footnote{Notice that errors of the first three data points are tightly correlated. So are the next two data points. Therefore effectively we have only three independent data points.} If we parameterize the differences into an overall amplitude $A$ through $R = AR_{\rm{Planck}}$, $A = 1.08 \pm 0.09$. \cite{Prat2019} measured the lensing ratios using galaxy position and lensing data from DES, and CMB lensing data from the SPT+\emph{Planck}. They found a best-fitting lensing ratio amplitude of $A = 1.1 \pm 0.1$. Our result of amplitude fitting is consistent with the previous work.

\begin{figure}
    \centering \includegraphics[width=\columnwidth]{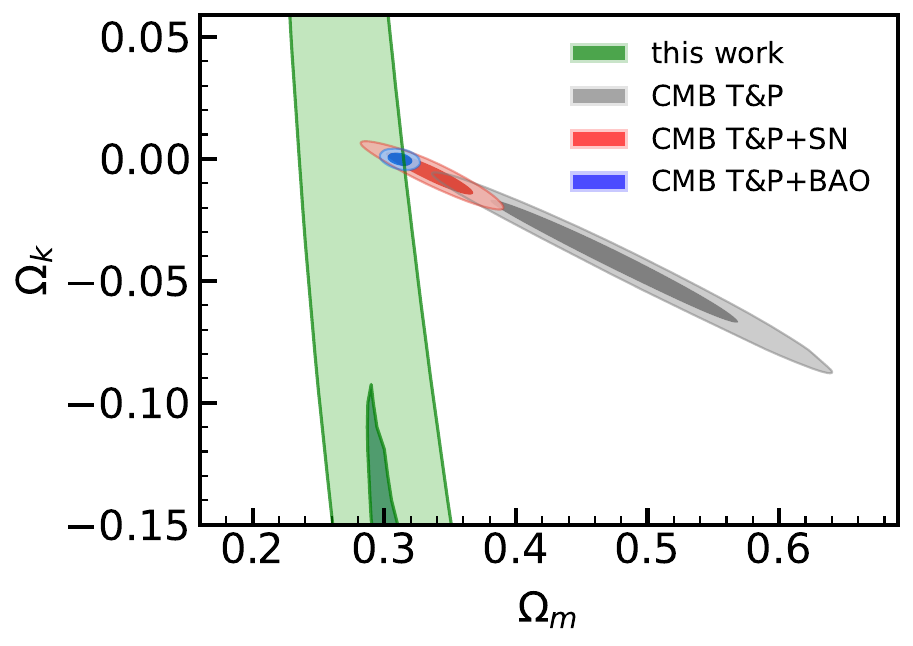}
    \caption{Cosmological constraint under the assumption of a model with a $w=-1$ cosmological constant with free curvature. The $\Omega_m - \Omega_k$ constraint uses the six measured lensing ratios (green), CMB (gray), CMB + SN (red), and CMB + BAO (blue). }
    \label{fig:Omegak_contour}
\end{figure}

It is beyond the scope of this paper to fully investigate the above issues. Instead, we briefly discuss two possibilities. One possibility is that the universe is non-flat. Fig.~\ref{fig:R_Omegak} explores this possibility by varying $\Omega_K$, while fixing all other cosmological parameters to the \emph{Planck} 2018 \citep{Planck2018parameters}. $R$ indeed varies with $\Omega_K$. However, the required modification of $\Omega_K$ ($\Omega_K\rightarrow -0.4$) is likely too dramatic to explain the data. Fig.~\ref{fig:Omegak_contour} shows our result of the $\Omega_m - \Omega_k$ constraint compared with eBOSS cosmology \citep{eBOSS2021}. Using six lensing ratios leads to a detection of $\Omega_m = 0.28_{-0.07}^{+0.05}$. The measured ratios do not provide a strong constraint on the cosmological curvature. 

\begin{figure}
    \centering
    \includegraphics[width=\columnwidth]{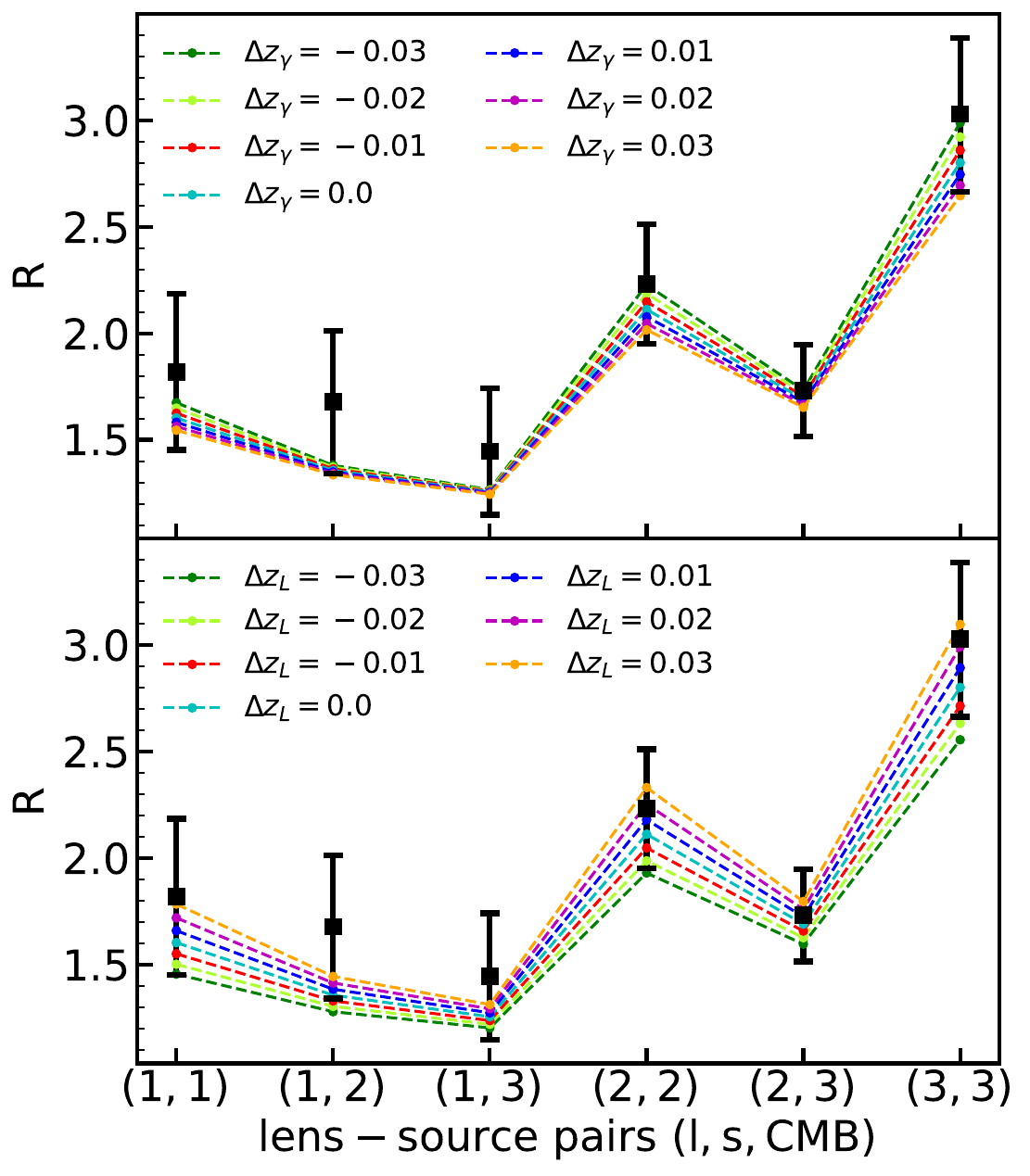}
    \caption{The impact of lens (shear source) redshift bias on the lensing ratios. }
    \label{fig:R_deltaz}
\end{figure}

Another possibility is the redshift uncertainties in the lens redshift and/or shear source redshift, as noticed in previous works (e.g. \citet{Das2009, Kitching2015, Prat2019}). Because of this, shear ratios can cross-check the redshift distributions. This has been done in the Sloan Digital Sky Survey \citep[SDSS,][]{Mandelbaum2005}, where both redshifts and multiplicative shear biases were tested for the first time. Afterwards, in the Kilo-Degree Survey \citep[KiDS,][]{Giblin2021, Heymans2012, Hildebrandt2017, Hildebrandt2020, Asgari2021} and in Dark Energy Survey \citep[DES,][]{Davis2017, DESY1_2018, DESY3_2021, DES_Y3_3x2pt} tested the redshift uncertainties are $\sim 0.01$. But DECaLS shape catalog we used in this analysis has not done this yet.

Redshift uncertainties can show as the r.m.s. $\sigma_z$ of the estimated $z^P$ or as a systematic bias $\Delta z$ \citep{Baxter2019, Prat2019}. 
(1) The mean $\sigma_z$ of tracer galaxies is $\sim$ 0.019, 0.034, 0.043 at redshift bin 0.1-0.3, 0.3-0.5, 0.5-0.7, respectively.  Doubling $\sigma_z$ only affects $R$ by less than 4\%.  Thus, $\sigma_z$ is not the major reason for the $\sim$ 10\% overestimation of ratios.  (2) $\Delta z$ changes the redshift distribution from $n^i(z)$ to  $n^i(z + \Delta z)$ (Eq.~\ref{eq:modeling the R}).
Fig.~\ref{fig:R_deltaz} shows the theoretical prediction of its impact. Positive $\Delta z_L$ for lens samples, or negative $\Delta z_\gamma$ for shear samples, are more consistent with the measured lensing ratio. $\Delta z\sim 0.01 - 0.03$ is sufficient to explain the found overestimation of $R$. Therefore the possible bias in the mean redshift should be included to correctly interpret the lensing ratio measurement.


\bibliography{sample631}{}
\bibliographystyle{aasjournal}



\end{document}